# Cascaded Multi-cycle terahertz driven ultrafast electron acceleration and manipulation

Dongfang Zhang[1]*, Moein Fakhari[1,2], Huseyin Cankaya[1,2], Anne-Laure Calendron[1], Nicholas H. Matlis[1] and Franz X. Kärtner[1,2]

**Affiliations:**

[1]Center for Free-Electron Laser Science, Deutsches Elektronen Synchrotron, Notkestrasse 85, 22607 Hamburg, Germany.

[2]Department of Physics and The Hamburg Centre for Ultrafast Imaging, University of Hamburg, Luruper Chaussee 149, 22761 Hamburg, Germany.

*To whom correspondence should be addressed. E-mail: dongfang.zhang@cfel.de

**Abstract**: Terahertz (THz)-based electron acceleration and manipulation has recently been shown to be feasible and to hold tremendous promise as a technology for the development of next-generation, compact electron sources. Previous work has concentrated on structures powered transversely by short, single-cycle THz pulses, with mm-scale, segmented interaction regions that are ideal for acceleration of electrons in the sub- to few-MeV range where electron velocities vary significantly. However, in order to extend this technology to the multi-MeV range, investigation of approaches supporting longer interaction lengths is needed. Here, we demonstrate first steps in electron acceleration and manipulation using dielectrically-lined waveguides powered by temporally long, narrowband, multi-cycle THz pulses that co-propagate with the electrons. This geometry offers centimeter-scale single-stage interaction lengths and offers the opportunity to further increase interaction lengths by cascading acceleration stages that recycle the THz energy and rephase the interaction. We prove the feasibility of THz-energy recycling for the first time by demonstrating acceleration, compression and focusing in two sequential Al$_2$O$_3$-based dielectric capillary stages powered by the same multi-cycle THz pulse. Since the multi-cycle energy achievable using laser-based sources is currently a limiting factor for the maximum electron acceleration, THz energy recycling provides a key enabling factor for reaching relativistic energies with existing sources.

**Introduction**

Femtosecond duration relativistic electron beams are in demand for a wide range of applications from ultrafast electron diffraction (UED) [1,2] and microscopy (UEM) [3], to ultrafast X-ray sources [4]. Microwaves in the radiofrequency (RF) regime (1–10 GHz) have been the conventional choice for accelerating electrons, but face significant challenges in meeting the request for future acceleration and manipulation of particle beams on the femtosecond time scale. The limited acceleration fields sustainable at low frequency also necessitate km-scale facilities to reach highly relativistic beams, making these devices highly exclusive and costly. Therefore, there is a strong motivation for exploring alternative solutions which are more compact, economic and adapted for pushing the resolution frontiers. Novel laser-based accelerator concepts provide intrinsic optical synchronization, allow scaling to smaller accelerator structures and generate substantially stronger fields for acceleration and beam manipulation. A prime example is laser-plasma accelerators (LPAs) [5,6], which can reach field gradients of more than 10 GV/m and have produced multi-GeV electron beams with percent level energy spread. However, LPAs need considerable laser development to operate at high repetition rates and are hampered by instabilities and difficulties in controlling beam parameters associated with the dynamical nature of the plasma. Dielectric laser accelerators (DLA) can reach GV/m accelerating fields [7,8], however present significant challenges in beam control and are so-far limited to low bunch charges in the single-electron range due to their sub-micron scale cross-sections.

Recently, there has been significant interest in the use of THz based particle acceleration and beam manipulation as a new approach with the potential to overcome certain limitations of RF accelerators and other laser-based concepts. In particular, the picosecond time-scale of laser-generated THz sources is expected to enable order-of-magnitude increases in sustainable field strength [9,10] to the GV/m [11–14] range, and increased precision in timing compared to RF sources while still supporting moderate charge at the picocoulomb level [15]. Proof-of-principle demonstrations with single-cycle THz pulses have resulted in multi-keV acceleration [16,17] and high-field manipulations of electron-bunch phase space [18–20] proving the feasibility of compact THz-based devices for future radiation sources or as components for boosting the performance of existing accelerators as well as a promising solution to produce high repetition, high energy ultrafast electron bunches, which are highly desired in ultrafast electron diffraction [21,22]. However, the interaction lengths in this configuration are limited by available single-cycle pulse energies to the millimeter range. A THz-driven inverse free-electron laser (IFEL) scheme has been proposed [23,24] that provides a potential solution to solve the dispersion, dephasing, and walk-off problems for relativistic electron using single-cycle THz [24].

Co-propagation designs powered by narrowband multi-cycle THz radiation have the potential to improve the interaction length as well as the symmetry of the field distribution. In particular, dielectrically-lined waveguides (DLWs) have attracted interest in the accelerator community for their versatility to accelerate, manipulate and characterize [25,26] electron beams. They require a simpler fabrication process using commercial solutions with percent-level precision [27] which enable tuning of the phase and group velocity of the THz wave inside the DLW. Proof-of-principle experiments were performed by Nanni *et al* [16] demonstrating the feasibility of THz-driven DLW accelerators. However, these experiments used single-cycle pulses of 10 µJ to modulate by 7 keV the energy of electrons injected at 60 keV. The interaction length was limited to 3 mm by severe dispersion of the THz pulse in the waveguide and by dephasing of the electrons due to the mismatch between the electron speed and THz phase velocity. Meanwhile the long injected electron bunch leads to not only acceleration but also deceleration at the same time. Dispersion, dephasing and temporal walk-off are three critical effects which must be managed to achieve long interaction lengths. Dispersion results from the frequency dependence of the THz phase velocity in the waveguide, and causes the pulse to temporally stretch and reduce in field strength; dephasing arises when the electrons slip from the accelerating to the decelerating phase of the field due to a difference between the electron velocity and the THz phase velocity; and "walk-off" arises when the electron bunch loses overlap with the THz pulse envelope due to a difference between the electron velocity and the THz group velocity. Dispersion can be mitigated by using THz sources of sufficiently narrow bandwidth which are now available with sub-mJ energies and sub-percent bandwidths [28], due to recent developments in THz generation using periodically poled lithium niobate (PPLN) [29] . Use of narrowband pulses can also lead to improved THz spatial mode quality [30], simplified beam transportation and improved coupling efficiency into the waveguides. Addressing dephasing and walk-off, however, requires development of structures and methodologies beyond the simple DLW concept. Tapering of the DLW properties has been proposed as one solution for addressing phase slippage [31]. However, this approach requires high injection energies and very tight tolerances on fabrication, and has so far not been demonstrated.

Here we explore, for the first time, use of narrow-band THz pulses to drive a DLW and introduce a new scheme based on cascading of DLWs which enables both re-phasing and re-timing of the electrons relative to the THz pulse, demonstrating re-use of the THz energy for the first time. Relative to previous work [16], we increase the interaction length by an order of magnitude to ~3 cm. Using only ~200 nJ of THz pulse energy, we accelerate 53 keV electrons by ~1.6 keV, which represents a ~10-fold improvement in energy transfer between the THz and the electrons, compression of the electron bunch by a factor of 10 down to ~150 fs (FWHM) and spatial focusing the electron beam by a factor of 2.5.

## Results

**Experiment Setup**

In the experimental setup [Fig.1], a 53 keV photo-triggered DC gun is used as an injector for the multi-stage DLW-based electron accelerator and manipulator which is powered by narrowband, multi-cycle THz pulses. The electron beam from the DLW device is analyzed by a segmented terahertz electron accelerator and manipulator (STEAM) device [20], which is used as a streak camera to measure the bunch length with a resolution of ~ 150 fs, and by a tunable electromagnetic dipole coupled with a micro-channel plate (MCP) which is used to measure the energy with a resolution of ~ 1 keV and deflection of the beam. Ultraviolet (UV) pulses for photoemission in the DC gun, multi-cycle THz pulses to drive the DLW device and single-cycle THz to drive the STEAM device are all created using a single infrared Yb:KYW laser system producing 4-mJ, 650-fs (FWHM), 1030-nm pulses at a 1 kHz repetition rate [32]. The UV pulses are generated by two successive stages of second harmonic generation (SHG), 50 ps (FWHM) long multi-cycle THz pulses are generated by intra-band difference frequency generation in a 5 mm long MgO:PPLN crystal [33] [Fig. 1 insert], and single-cycle THz pulses are generated via the tilted-pulse-front method in a $LiNbO_3$ prism [11]. The linearly polarized multi-cycle THz beam, which has 13 cycles and a center frequency of 0.26 THz, is then converted to a radially polarized beam via a segmented waveplate before being coupled into the DLW device collinearly to the electron motion using an off-axis-parabolic mirror and horn structure that concentrated the THz fields into the interaction zone.

The DLW design consists of a cylindrical copper waveguide of diameter 790 µm and length 30 mm with a dielectric layer of alumina ($Al_2O_3$, THz refractive index n=3.25, see supplementary) with a wall thickness of 140 µm. The waveguide supports a travelling transverse-magnetic waveguide mode ($TM_{01}$ mode), the axial component of which provides the longitudinal field for acceleration and deceleration [Fig. 2]. The dimensions and index of the dielectric material are chosen to according to Fig. 2 that provides a phase velocity of ~0.43c at 0.26 THz, which matches the initial velocity of the electrons and optimizes the device for electron acceleration and manipulation functions. Alumina is also desirable as a dielectric material because of its transparency and high damage threshold. The dimensions and index of the DLW result in a group velocity of the THz pulse that is ~0.25c, which is significantly slower than the electron velocity and leads to a strong walk-off effect. The diameter of the vacuum core of the DLW is 510 µm which represents the clear aperture for the electron propagation.

Conical horns are used to couple THz energy into and out of the waveguides. In the acceleration mode, 3.5 mJ of the Yb:KYW laser pulse energy is used for multi-cycle THz generation, resulting in 200 nJ THz energy injected into the DLW via the horn coupler. For compression and focusing, 0.5-mJ laser pulses are used for multi-cycle THz generation and the rest (3 mJ) is used for single-cycle THz generation. The single-cycle pulses centered at 300 GHz, which have an energy of 6 µJ, are injected into the STEAM streak camera [20] which has a time resolution of ~150 fs.

**Operation**

The function of the device is selected by tuning the relative delay between the THz pulse and the electron bunch which determines the phase of the THz waveform that the electron bunch experiences. There are four key phase points within the waveform corresponding to distinct behaviors: the positive and negative crests of the waves, where the field gradient is minimized, and the positive and negative "zero-crossings" of the field, where the field gradient is maximized. The positive and negative crests correspond to deceleration and acceleration of the electron bunch, respectively, but leave the bunch spatial and temporal dimensions unchanged. At the zero-crossings, by contrast, the average energy and transverse position of the bunch are unchanged, but the bunch experiences a combination of spatial and temporal reshaping. At the positive zero-crossing, where the field gradient is positive, the bunch becomes stretched in time but focused in space, while at the negative zero-crossing, the bunch is temporally compressed but expands in space. As described by the Panofsky–Wenzel theorem [34], the longitudinal gradient of the transverse force is proportional to the transverse gradient of the longitudinal force.

To optimize the energy gain of a single DLW stage, the interaction length in the accelerating phase must be maximized, and decelerating phase must be avoided. The interaction length can be maximized by initially setting the phase velocity to be larger than the injected electron speed by an amount just enough that electrons can gain energy and start to overtake the phase velocity of the THz wave. To avoid the decelerating phase, there are two methods that can be used. The first method relies on coupling the THz wave out of the DLW before electrons slip into the decelerating half-cycle. Another way is to take advantage of walk-off between the THz pulse and the electron bunch by designing the THz to be short enough that electrons enter the decelerating half-cycle only after they have overtaken the envelope of the THz field. The second method ensures much cleaner acceleration, because the electrons do not interact with the less homogenous fields in the coupler region.

For compression and focusing, the speed of the electron bunch should be matched to the phase velocity of THz pulse inside the waveguide since neither will change during the interaction. The interaction length is then only limited by the THz pulse duration and the rate of walk-off. In order to minimize unwanted effects at the entrance and exit boundaries, it is desirable that the complete interaction occur inside the DLW. As long as the THz group velocity is less than the electron velocity, a clean THz-electron interaction can be achieved by choosing a waveguide of sufficient length so that the electrons have overtaken the THz pulse before exiting.

Compression of the electron bunch is based on "velocity bunching" [35]. In this technique, the electron bunches are placed at the negative zero-crossing so that the tail experiences acceleration and the head experiences deceleration, but the bunch sees no average energy gain. The imparted momentum gradient then leads to compression of the bunch during propagation. For stretching, the bunch is placed at the positive zero-crossing and experiences the opposite gradient.

The focusing and defocusing of the bunch which also occurs at the zero crossings are due to the transverse electric fields. These fields are partially canceled by the transverse-deflection force from the magnetic field of the travelling THz.

In a circular waveguide, the net transverse force on an electron which is moving with velocity $v_e$ parallel to the z-component of the electric field [36] can be derived:

$$F_\perp = F_e + F_m = q(E_r - v_e B_\varphi) \propto q\left(1 - \frac{v_e v_{ph}}{c^2}\right)$$

where, c is the speed of light, $v_e$ is the elctron velocity, $v_{ph}$ is the phase velocity of the electric field, $q$ is the charge of the electron, $F_e$ and $F_m$ are the forces of the electric and magnetic components respectively. As can be seen from the above equation, for the relativistic electrons which velocity is close to *c*, the net transverse field tends to vanish. However, for non-relativistic electrons or when $v_{ph}$ or $v_e$ is less than *c* there would be a net deflecting force acting on the electrons.

**Single DLW Results**

Figure 3 shows the acceleration results from a single DLW. As the delay between the electrons and the THz is varied, a cross-correlation between the THz and electron can be seen [Fig. 3(e)]. A maximum energy shift of around 1 keV is observed using 200 nJ multi-cycle THz pulse energy [Fig. 3(a)]. Due to the low THz pulse energy, the energy change is quite small. However, the energy gain per unit of THz pulse energy is over 7x higher than in the previous results, demonstrating the benefit of increased interaction lengths in the co-propagating geometry. We achieve a THz-electron interaction distance of ~15 mm which is more than 5 times longer than that previously reported using single-cycle THz pulses [16,20]. Compression of the bunch is shown in Fig. 3(b). At maximum compression the electron bunch duration was reduced by a factor of 8 to a full-width half maximum of 180 fs (FWHM), which is in agreement with the simulation performed with the particle-tracking program ASTRA [37]. A clear difference between compressed and uncompressed beam can be seen by the streaking deflectograms generated by plotting a lineout of the spatial charge distribution along the streaking dimension as a function of delay relative to the THz field [Fig. 3c]. Focusing of the electron beam is shown in Fig. 3(d). At maximum focusing, the beam radius is reduced by a factor of 2.5 and the peak intensity increased by ~6 times. The quality of the focused beam is similar to that produced by the conventional electromagnetic lens in front of the DLW.

**Cascaded DLW Results**

In the single DLW experiments, limited by the available THz energy the initial velocity of the electrons is designed to match the phase velocity of the THz and the absolute average energy gain is sufficiently small, even for the acceleration experiments, that phase slippage is not the limiting factor. The limiting factor for the interaction length, in fact, is walk-off due to the large difference between the electron and pulse-envelope velocities. However, because the relative speed of the THz pulse and the electron bunch reverses outside the DLW, it is possible to recycle the THz pulses for a second interaction. As illustrated in Fig. 4(a), inside DLW I, the injected electrons travel faster than the envelope of the THz pulse, and therefore overtake it. A horn coupler and a THz lens with a hole in the center are then used to couple THz into free space and collimate it while allowing the electrons to pass through. In vacuum, the THz pulse has a higher group velocity than the electron bunch, and thus catches up and overtake the electrons. During this propagation, the THz field has a negligible effect on the electrons due to the large size and resultant low field strength of the collimated THz beam. The THz and electrons can then be coupled into a second DLW for a second interaction. The timing between the electrons and THz in the second DLW is set by controlling the distance between DLWs I and II using a motorized stage. Fine adjustments of the DLW II longitudinal position can also be used to select its operation mode. For example, to switch from focusing to defocusing requires shifting of the electrons by 180 degrees in phase relative to the THz waveform. Taking into account the speed of both the THz pulse and electron bunch, the required one cycle in position of DLW II is: $v_e*\lambda/(c-v_e)=0.856$ mm, which is less than half of the THz wavelength ($\lambda=1.153$ mm at 260 GHz).

The results of the cascaded acceleration and compression experiments are shown in Figure 4. By recycling the THz from DLW I, the electrons are further accelerated to 54.6 keV, nearly doubling the energy gain. When electrons are acceleration using fast varying THz field, the bunch length needs to be much shorter than the duration of a THz half-cycle. This is related with the variation of the electric field experienced by different electrons within the bunch arriving at different times. In general, larger field gradient and longer bunch length will lead to larger energy spreads [38]. For DLW I, the injected bunch length is ~ 300 fs (FWHM). For DLW II slight velocity bunching is performed in DLW I to enable still short electron bunch when injected into DLW II. Due to the small energy gain and limited energy resolution of the dipole spectrometer, no obvious energy spread is observed. However, this can be especially important for high energy acceleration, where much shorter bunch length is desired to achieve low energy spread [38]. Similarly, the cascaded compression results in a reduction of the bunch duration down to 150 fs (FWHM), which is the limit of the streaker's time resolution. As can be seen from the simulations in Fig 4(e), the actual bunch length is most likely shorter than was measured. If the longitudinal focus is too long, a single DLW I cannot compress the beam tightly. While using only DLW II, the electron bunch becomes too long that it exceeds the compression range of DLW II. Two DLWs help maintain the beam quality and tight longitudinal focus at the desired position. Due to the recycling of the THz, the required energy for compression is reduced by around 10 %. By contrast, the observed improvement in focusing is not very large, and is likely limited by the transverse emittance of the injected beam and long distance to the detector. Based on the acceleration results and the electron energy gain derived from simulations [Fig. 4(d), top], the THz coupling from the first to the second stage is around 10 %. This poor figure can be greatly improved by optimizing the THz telescope and horn coupler design, for example, adding coating on the lenses, using corrugation and curvature on the horn, and implementing coupler to match the impedance.

We have also simulated THz acceleration with 35-cycle THz pulses, 200 µJ energy and a centre frequency of 300 GHz, which is achievable with state-of-the-art THz technology [28]. With around 100 MV/m longitudinal electric fields achieved inside the waveguides, which is well below the field breakdown thresholds in dielectric structures [39], 55 keV electrons can be accelerated to around 300 keV in the first stage and beyond 1 MeV in the second stage, with phase-velocity settings of 64 % and 89 % of the vacuum speed of light for DLW I and DLW II, respectively. Due to the quick phase slippage under high-field-gradient conditions, DLW I must be kept very short or the timing between electron and THz pulse must be controlled in order to limit the interaction length and prevent dephasing. As can be seen from Fig. 4(d), the interaction length is much shorter while accelerating with 100 MV/m THz-fields when compared with lower field gradient. Higher fields lead to shorter interaction lengths since the electron speed changes much faster. As the electron beam reaches relativistic speeds, it requires more drift time in free space for the THz pulse to overtake the electron bunch and thus a longer separation between DLWs I and II and longer travel distances for DLW II to switch between different operation modes. For example, it will take ~1690 mm for a 100 mm long THz pulse to overtake a 1 MeV (94%c) electron beam and the relative cycle in DLW II is ~16 mm. As the electron beam accelerates over successive stages, the interaction length must also increase, as can be seen in DLW II [Fig. 4(d), bottom].

**Summary**

We demonstrate acceleration, temporal compression and spatial focusing of fC charge electron bunches by using a cascaded DLW device powered with multi-cycle THz pulses. DLWs provide homogeneous spatial field distributions and are easy to implement. Taking advantage of the large group velocity difference, we are able to achieve the manipulation of electrons over centimeter scale interaction lengths, which can be increased to even longer interaction distances by further recycling of the THz pulses via the cascaded manipulation scheme, hence lower THz energy required for the desired energy gain. This is especially important as the efficiency of current laser-based technologies for THz generation is still rather low (<1%), which requires J-level laser systems for MeV level electron acceleration. This cascading scheme will greatly lower the demand on the required laser system for electron acceleration in the non-relativistic regime such as in the development of THz injector guns and ultrafast electron diffraction sources.

Currently, development of laser-based narrowband THz generation is very intense and has resulted in demonstration of 0.6 mJ, sub-percent-bandwidth THz pulses [28] which are ideal for co-propagating THz-based acceleration. Numerical calculations predict that with our scheme, only 200 µJ of THz energy is needed to reach the MeV threshold, which would represent a milestone achievement in THz acceleration. Schemes for reaching multi-mJ energies have also been described, are numerically analyzed, and are in process of being tested [40,41]. It is therefore timely and pertinent to develop appropriate techniques for exploiting these upcoming powerful THz sources that can power multi-MeV electron sources that will truly be disruptive.

Multiple functions can be implemented using a single multi-cycle THz pulse by proper phasing of each stage. The physics insight provided by the cascaded manipulation scheme paves the way for further practical implementation of THz-driven acceleration in high-energy physics, ultrafast electron diffraction, and tabletop x-ray sources.

**Acknowledgments:** The authors would like to thank Arya Fallahi for helpful discussions and the technical support of Thomas Tilp and Matthias Schust for fabrication of the DLW-device used in this work. We would also like to acknowledge manufacturing of the magnetic steer for energy measurement by the group of Dr. Ralph Assmann at DESY. This work has been supported by the European Research Council under the European Union's Seventh Framework Programme (FP7/2007-2013) through the Synergy Grant AXSIS (609920), the Cluster of Excellence 'Advanced Imaging of Matter' of the Deutsche Forschungsgemeinschaft (DFG) - EXC 2056 - project ID 390715994 and the accelerator on a chip program (ACHIP) funded by the Gordon and Betty Moore foundation (GBMF4744).

**Figures**

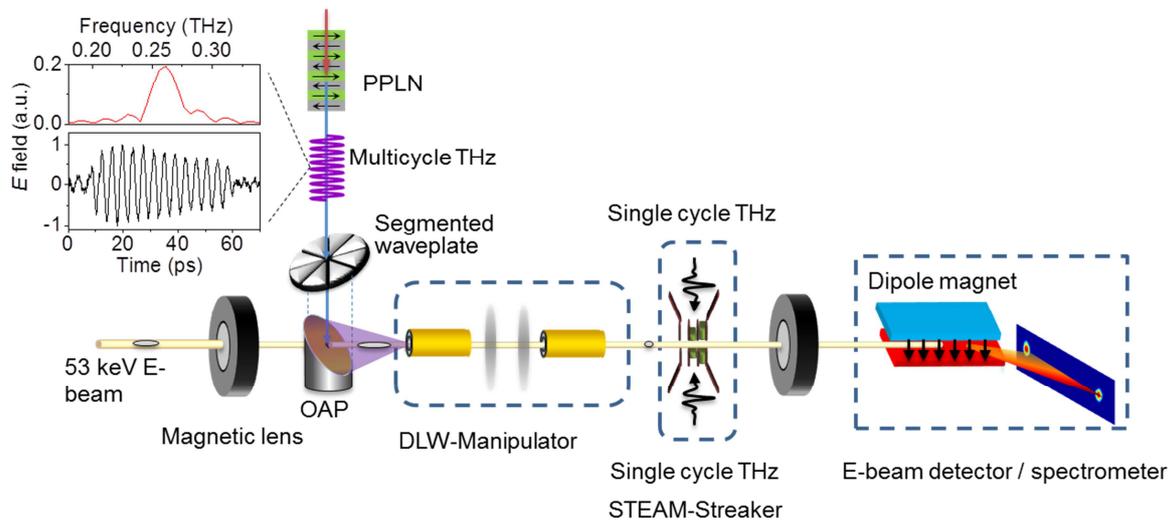

FIG. 1 Experimental setup. A small fraction of the infrared optical beam is converted to 257 nm through fourth harmonic generation. The 257-nm laser pulse is directed onto a gold photocathode generating electron pulses, which are accelerated to 53 keV by the DC electric field with around 1 fC charge. The same infrared laser also drives a multi-cycle THz generation stage and two single-cycle THz stages for the DLW-manipulator and the STEAM-streaker, respectively.

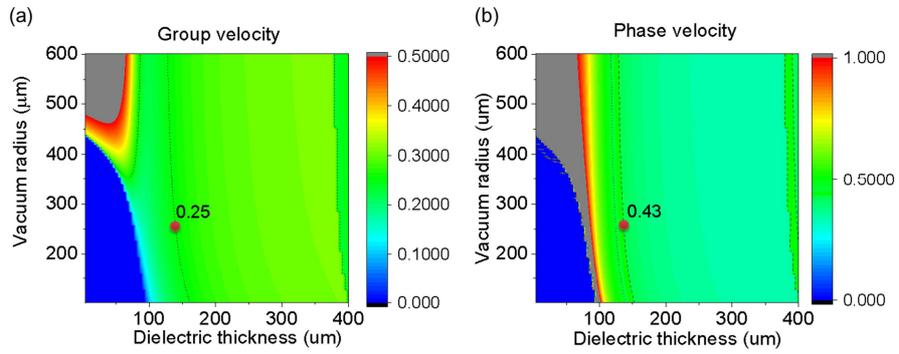

FIG. 2 $TM_{01}$ - mode parameters of DLW. Group (a) and phase (b) velocity of the $TM_{01}$ mode at 260 GHz in a circular copper waveguide with dielectric loading. The red dots represent the design point for electron beam manipulation with 250 µm vacuum radius and 140 µm dielectric thickness.

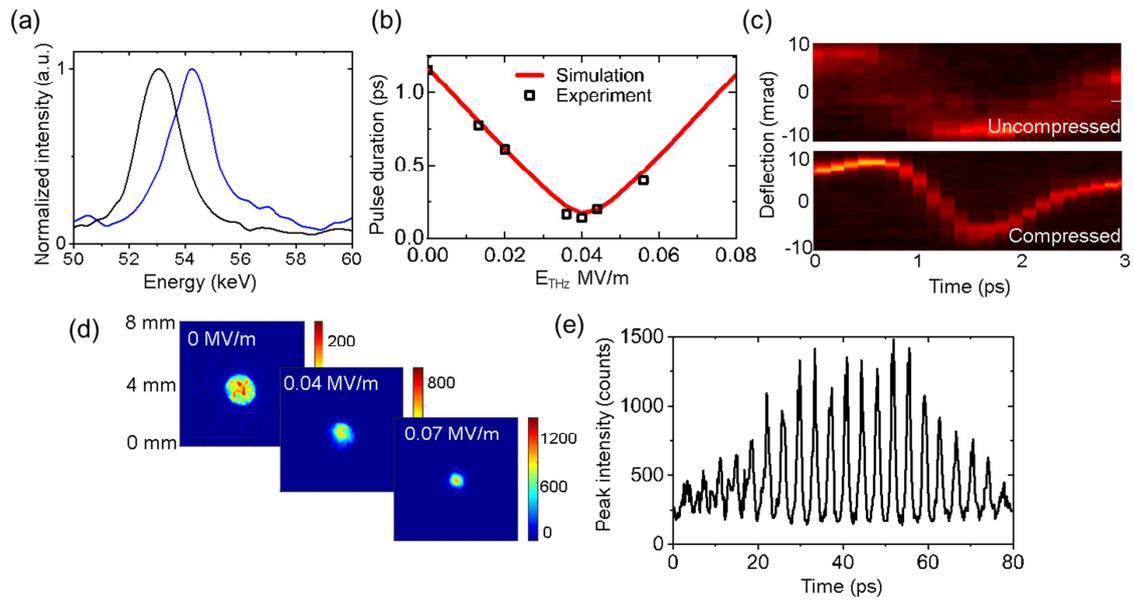

FIG. 3 DLW based electron manipulation. (a) Measured electron energy spectra for input beam (black curve) and accelerated beam (blue curve) in the acceleration mode. (b) Measured (hollow square) and simulated (red line) electron bunch length as a function of applied THz field in compression mode. (c) Time-dependent deflection diagrams measured by varying the delay between the arrival time of the electron bunch and the deflecting THz pulse for initially uncompressed and compressed electron bunches. (d) Electron beam diameter as a function of the THz field strength in focusing condition. (e) Electron peak intensity change as a function of THz-electron delay.

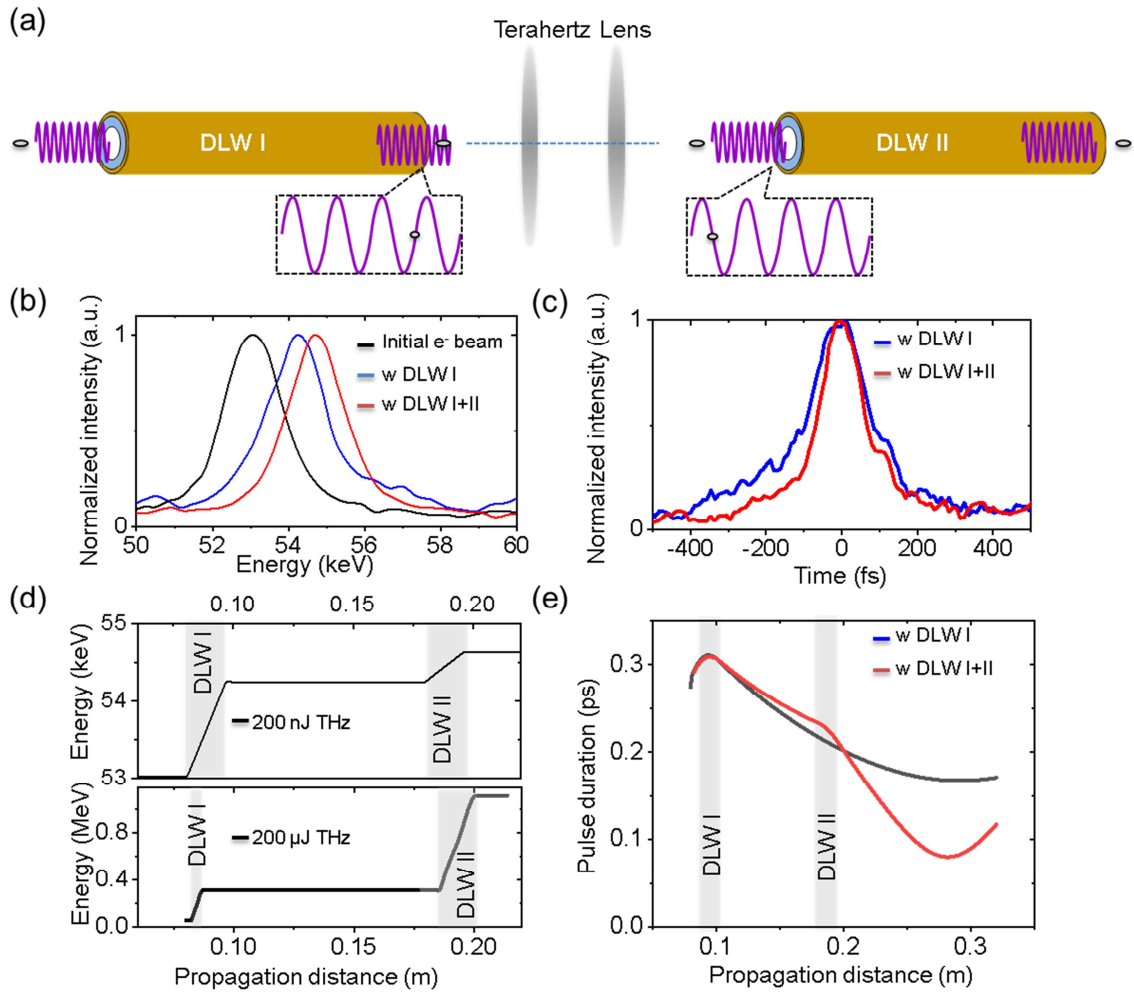

FIG. 4 Cascaded manipulation. (a) Schematic of the cascaded acceleration. Due to the low group velocity of the accelerating THz field inside the DLW, after THz-electron interaction, the electrons will overtake the THz wave at the exit of DLW I. In vacuum, the THz pulse will catch up and overtake the electron bunch again before both enter DLW II for a second interaction, i.e. acceleration stage. (b) Measured electron energy spectra for input beam (black curve), single stage accelerated beam (blue curve), and cascaded electron acceleration (red curve). **c,** Measured pulse duration via the single stage compression (blue curve) and two stage compression (red curve). (d) Top: simulation of experimental condition with two stage cascaded acceleration that couples 200 nJ THz pulses with ~0.11 MV/m longitudinal electric fields into the DLW I. The THz coupling efficiency from the 1st to 2nd stage is 10 %. Bottom: simulation of two stage cascaded acceleration with 200 µJ THz pulse, ~100 MV/m field gradients inside the dielectric waveguide, and 90 % THz coupling. (e) Simulation of electron pulse duration (FWHM) during propagation with single DLW stage and two DLW stages. The coupling of THz into the second stage is assumed to be 10 %.